\documentstyle[aps]{revtex} 

\newcommand{\twooneplaq}{\setlength{\unitlength}{.5cm}
   \raisebox{-.2cm}{
   \begin{picture}(2.2,1.2)(-1.1,-.6)
   \put(-1,-.5){\line(1,0){2}}
   \put(-1,.5){\line(1,0){2}}
   \put(-1,-.5){\line(0,1){1}}
   \put(1,-.5){\line(0,1){1}}
   \put(-1,-0.5){\circle*{.2}}
   \put(-1.5,-0.9){$x$}
   \put(-0.05,-1.2){$\mu$}
   \put(-1.5,-0.05){$\nu$}
   \end{picture}}}
\newcommand{\ltwooneplaq}{\setlength{\unitlength}{.5cm}
   \raisebox{-.2cm}{
   \begin{picture}(2.2,1.2)(-1.1,-.6)
   \put(-.5,-1){\line(1,0){1}}
   \put(-.5,1){\line(1,0){1}}
   \put(-.5,-1){\line(0,1){2}}
   \put(.5,-1){\line(0,1){2}}
   \put(-0.5,-1){\circle*{.2}}
   \put(-0.9,-1.5){$x$}
   \put(-1.2,-0.05){$\nu$}
   \put(-0.05,-1.5){$\mu$}
   \end{picture}}}

\newcommand{\be}{\begin{equation}}
\newcommand{\ee}{\end{equation}}
\newcommand{\bea}{\begin{eqnarray}}
\newcommand{\eea}{\end{eqnarray}}

\def\tint{\tau_{int}}

\begin{document}

\draft

\title{\bf Noisy Monte Carlo revisited}

\author{T. Bakeyev$^1$ and Ph. de Forcrand$^{2,3}$}   
\address{$^1$Joint Inst. for Nuclear Research, 141980 Dubna, Russia }
\address{$^2$Inst. f\"ur Theoretische Physik, ETH-H\"onggerberg, CH-8093 Z\"urich, Switzerland}
\address{$^3$CERN, Theory Division, CH-1211 Geneva 23, Switzerland}
\date{\today} 
\maketitle

\begin{abstract}
We present an exact Monte Carlo algorithm designed to sample theories where
the energy is a sum of many couplings of decreasing strength.
Our algorithm, simplified from that of L. Lin et al. \cite{KFLiu}, avoids
the computation of almost all non-leading terms. We illustrate its use
by simulating $SU(2)$ lattice gauge theory with a 5-loop action, 
and discuss further applications to full QCD.
\end{abstract}


\section{Introduction} 

When sampling by Monte Carlo the partition function 
$Z = \int \prod dU ~~ e^{-H(\{U\})}$,
the most common algorithm is that of Metropolis \cite{Metropolis}:
at each step, starting from the current configuration $\{U\}$, a candidate
configuration $\{U'\}$ is proposed, and it is accepted with probability
\be
P_{acc} = \rm{min}(1, e^{- (H(\{U'\}) - H(\{U\}))} )  \ . 
\label{acc}
\ee
This acceptance test is realized by comparing the right-hand side of (\ref{acc})
to a random number uniformly distributed in $[0,1]$. This seems like a waste of
information: why compute $H(\{U'\})$ {\em exactly}, then compare it with a random
number? It should be sufficient to {\em estimate} it. Indeed, this logical 
proposition has been studied several times \cite{KFLiu,KK,BK}. Two difficulties
have been identified, both caused by the non-linear relationship between the
energy $H$ and the probability $\propto e^{-H}$: 
$(i)$ what is needed is an unbiased estimate of $e^{-H}$, which must be obtained
from unbiased estimate(s) of $H$; 
$(ii)$ to be interpreted as a probability, the noisy estimator of $P_{acc}$ must
be bounded, and in particular stay positive.
Difficulty $(i)$ was overcome in \cite{BK}, which however showed that violations
of $(ii)$ caused intolerable systematic errors unless the amount of noise in the
estimate of $H$ was minuscule. Difficulty $(ii)$ was overcome in \cite{KFLiu},
which showed that exact results could be obtained even in the presence of a large
amount of noise in the estimate of $H$. Ref.\cite{KFLiu}, however, introduces
an infinite number of auxiliary variables, so that it may take an infinite amount
of work to bring these auxiliary variables to equilibrium. And tests of the 
method are performed on a toy model with 5 degrees of freedom only, whose 
relevance may be questioned. Here, we simplify the method of \cite{KFLiu},
by introducing only 1 auxiliary variable per term in $H$. Moreover, we separate
$H$ into a leading part to be calculated exactly, and a sum of small correction
terms, which we treat stochastically. This separation
is essential: because stochastic estimates are used for correction terms only,
large amounts of noise can be tolerated. As a consequence, our algorithm is
a very efficient approach to the simulation of complicated Hamiltonians.

Consider a generic Hamiltonian of the type
\be
H = \sum_{k=0}^m c_k W_k
\label{Ham}
\ee
where as $k$ increases, $|c_k|$ decreases and the successive terms $W_k$ typically become less and less local.
For instance, in a spin model $\{\vec{\sigma_i}\}$, $W_0$ would be the 
nearest-neighbour interaction $\sum_{<ij>} \vec{\sigma_i} . \vec{\sigma_j}$,
$W_1$ would represent next-nearest-neighbour interactions, etc... Here, we will
illustrate our method for lattice gauge theory. In that context, $W_k$ are
the traces of Wilson loops of increasing size: 
$W_0 = \sum_{x,\mu,\nu} Tr \prod_4 U$
around elementary plaquettes, $W_{1,2,3}$ correspond to different geometries of
6-link loops, etc...
It is often the case that one would like to study a Hamiltonian of type
(\ref{Ham}) resulting from an expansion, be it perturbative \cite{Symanzik}, non-perturbative
\cite{Lepage}, or based on the fixed point of a renormalization group 
transformation \cite{Hasenfratz}. In all these situations, the expansion is 
truncated to a maximal order $m$ dictated by technical reasons. 
As $k$ increases in (\ref{Ham}), the number of geometrically equivalent terms
grouped into $W_k$ increases exponentially: in a spin model on a hypercubic
lattice in $d$ dimensions, each spin has $2d$ nearest-neighbours (these interactions
are grouped into $W_0$), $d(d-1)/2$ next-nearest neighbours (grouped into $W_1$),
$d(d-1)(d-2)/6$ 3rd-neighbours, etc... This combinatoric explosion normally makes
the simulation of extended Hamiltonians prohibitively expensive. This is the
reason for a truncation to very low $m$, often taken to be 1 or 2.
However, in most cases the couplings $c_k$ in (\ref{Ham}) decrease exponentially
with $k$, so that the overall Hamiltonian is dominated by $W_0$, with small
corrections. In lattice field theory, this is actually required if the Hamiltonian
is to make sense and tend to a local operator as the continuum limit
is approached. By making use of stochastic methods to estimate the
correction terms $W_k, k \ge 1$, we aim at postponing the combinatoric explosion
of the simulation costs incurred when including higher terms $W_k$. This opens
the possibility of studying numerically much more complicated Hamiltonians
including higher-order correction terms. In lattice field theory, these
correction terms are crucial to suppress discretization errors, and form the
building blocks of so-called ``improvement'' strategies. Also the inclusion of
higher-order terms can be very useful in the approaches to the fermion
determinant simulations involving the loop expansion \cite{Forcrand}.

We present our method in Section II, and illustrate it in Section
III with simulations of a 5-loop perturbatively improved action 
for $SU(2)$ lattice theory. We conclude with prospective 
applications of our method, in particular for dynamical fermion QCD simulations.

\section{Noisy Monte Carlo: the method}

Given the Hamiltonian (\ref{Ham}) let us suppose that the terms $c_kW_k$
are nonpositive starting from $k=1$:
\be k\ge 1 :\quad c_k W_k(U)\le 0 ~~ \forall U \label{S11} \ee
This can be easily arranged by adding to each term of the Hamiltonian a 
nonessential constant. Here $U$ are the fields of the model under consideration. 
The key idea of the method is to estimate the contribution of the terms
$W_k(U),\; k\ge 1$ stochastically by introducing auxiliary fields.
This will lead to a significant reduction of computational effort if
the coefficients $c_k,\; k\ge 1$ are small enough. In all cases, the algorithm 
remains exact.

We introduce auxiliary fields $\sigma_k,\; k\ge 1$ (associated with
the terms $W_k$), which can take two values: 0 and 1. Using the identity
\be
a+b=\sum_{\sigma=0,1} [a*\delta_{\sigma,0}+b*\delta_{\sigma,1}]
\ee
we represent the probability $e^{-H}$ in the form
\be 
e^{-H}=P_0[U] *  P_1[U,\sigma]
\label{Intr} \ee
where
\be P_0[U]=e^{-c_0W_0(U)}\quad ;\quad  
P_1[U,\sigma]=\prod_{k=1}^{m}
\sum_{\sigma_k=0,1}[\delta_{\sigma_k,0}+\delta_{\sigma_k,1}(e^{-c_kW_k(U)}-1)]
\label{sigpr} \ee
The r.h.s. of (\ref{Intr}) can be interpreted as the joint probability distribution for the
original fields of the model and the new $\sigma$ fields. Due to the inequalities 
(\ref{S11}) this distribution is well defined: $P_1[U,\sigma]\ge 0 \;\;\;\forall \{U,\sigma\}$, and the probabilities for $\sigma_k$
to take value 0 or 1 when the $U$ fields are fixed lie in the interval [0,1]:
\be 
p_{\sigma_k=0}=e^{c_kW_k(U)}\quad ;\quad p_{\sigma_k=1}=1-e^{c_kW_k(U)}
\label{s13} \ee
This means that our algorithm has no probability bound violations, which plagued  
previous attempts to construct an efficient stochastic algorithm \cite{KK,BK}.

One can easily see why the introduction of auxiliary $\sigma$ fields can be useful.
Starting from the current $\{ U_1,\sigma\}$ configuration, a candidate configuration
$\{ U_2\}$ distributed with the weight $P_0[U_2]$ is proposed, and accepted with
probability 
\be
P_{acc} = \rm{min}\Bigl( 1, \frac{P_1[U_2,\sigma]}{P_1[U_1,\sigma]} \Bigr) =
 \rm{min}\Bigl( 1, \prod_{k\; :\;\sigma_k=1}\frac{e^{-c_kW_k(U_2)}-1}{e^{-c_kW_k(U_1)}-1}
\Bigr) 
\label{pacc} \ee
Since the terms $c_k W_k(U)$ contribute in $P_{acc}$ only if $\sigma_k=1$, the amount of 
computational work
is greatly reduced if the configurations with $\sigma_k=0$ are dominating. That is certainly 
the case when the absolute values of the coupling coefficients $|c_k|$ are small: 
the probabilities for $\sigma_k$ to be unity, 
averaged over $\{ U\} $ configurations, are negligible then.
Indeed, to leading order in $c_k$ the average probability $p_{\sigma_k=1}$ from 
eq.(\ref{s13}) can be written as
\be \langle p_{\sigma_k=1}\rangle  \;\; \approx \; -c_k \langle W_k(U)\rangle  \;\; \approx \; 0\quad  {\it if}\quad c_k\approx 0
\label{klm} \ee
Expression (\ref{klm}) also suggests that one should try to make $|\langle W_k\rangle |$ as small as
possible, using the freedom one has to shift $W_k$ by a constant. This goal
should remain compatible however with inequalities (\ref{S11}); otherwise, probability
bound violations will appear for $p_{\sigma_k}$ and $P_{acc}$ in eqs.(\ref{s13},\ref{pacc}).

Actually, the violation of conditions (\ref{S11}) is not completely forbidden. As it was pointed in Ref.
\cite{KFLiu}, one can address the problem of the lower probability-bound violations by redefining the
measure. If the distribution $P_1[U,\sigma ]$ in (\ref{Intr}) can be negative for some
configurations $\{ U,\sigma\}$, one can effectively
simulate with the probability distribution $P_0[U]*\Bigl|P_1[U,\sigma ]\Bigr|$ instead 
and include the sign $sgn (P_1)$ into the observable expectation value:
\be
\langle O\rangle =\frac{\langle O \; sgn(P_1)\rangle _{||}}{\langle sgn (P_1)\rangle _{||}}
\label{pviol}\ee
where by $\langle \rangle _{||}$ we denote the averages with respect to distribution 
$P_0[U] *  \Bigl|P_1[U,\sigma ]\Bigr|$.
Sometimes the admission of very rare sign violations can substantially decrease the
probability $p_{\sigma_k=1}$. However one should be very careful in using this trick:
as the volume of the system increases, one needs an exponential growth of statistics to estimate
$\langle sgn (P_1)\rangle _{||}$ within the same accuracy. In the following we shall always assume
fulfillment of the inequalities  (\ref{S11}).

After updating the $U$ fields one should also update the $\sigma$ fields to preserve ergodicity.
This requires the calculation of probabilities (\ref{s13}). 
At this point the reader might say: "OK, one saves computational effort by not calculating 
some terms $W_k$ in expression (\ref{pacc}) while estimating $P_{acc}$. Nevertheless,
one must calculate these terms when updating the $\sigma$ fields! So does one gain anything in the end?"
The answer is 'yes' for the following two reasons:
\begin{itemize}
\item 
The terms $W_k$ for which it is reasonable to use the stochastic estimation 
usually couple many degrees of freedom (this is due to the usual nonlocality of weakly coupled
terms, which serve as corrections to more local leading terms in the Hamiltonian). If one 
uses usual local algorithms (without introducing stochastic 
$\sigma$ variables), one should estimate the term $W_k$ each time one updates
a degree of freedom which it couples. Contrary to that, if one uses Noisy Monte Carlo, the probabilities 
(\ref{s13}) should be calculated only once per $\sigma$ update. 
\item
The variables $\sigma_k$ can be refreshed
infrequently, the more so as the associated coupling $c_k$ gets smaller. 
It will be demonstrated in the next Section on a particular example. This slow dynamics
of the auxiliary $\sigma$ fields does not imply slow dynamics of the 
physically relevant $U$ fields.
\end{itemize}

Up to now we were quite generic, showing that the Noisy Monte Carlo (NMC) method can be
potentially very effective for the variety of theories where the energy (\ref{Ham}) is a sum of couplings
of decreasing strength. In the next section we illustrate these ideas on a particular 
example: a 5-loop perturbatively improved SU(2) Yang-Mills model.

\section{5-loop SU(2) gauge theory}

We consider a 5-loop SU(2) gauge action in $4d$:
\bea 
S\hspace{-.3cm} \;\; &=&\;\; \hspace{-.2cm}  \sum_{i=1}^5 c_i \
{1 \over {m_i^2 n_i^2}}
\ S_{m_i,n_i}
\label{5Li}
\eea
where the indices $(m_i,n_i) = (1,1),(2,2),(1,2),(1,3),(3,3)$ 
for $i=1,\ldots , 5$ denote the planar, fundamental loops of size $m\times n$
\begin{eqnarray}
S_{m_i,n_i}\hspace{-.3cm}\;\; &=& \;\;\hspace{-.2cm}\sum_{x,\mu,\nu} 
\left( -2*{\rm sgn\; (c_i)} -\frac{{\rm Tr}}{2}\
 ( \   \twooneplaq + \hspace{-.1cm}\ltwooneplaq\ \hspace{-.1cm}
) \right) \label{act1}
\end{eqnarray}
The Gibbs factor is $\ {\rm exp}(- \frac{\beta}{2} S)$. Note that in eq.(\ref{act1})
we have arranged the constant term $\; -2*sgn\; (c_i)$ to ensure the condition 
(\ref{S11}) for elementary action terms corresponding to each loop:
\bea
~~ \forall \{ i,\mu,\nu,x,U\}  : \;\;
S_{i,\mu,\nu,x}\equiv
 {1 \over {m_i^2 n_i^2}} c_i 
\left( -{\rm sgn\; (c_i)} -\frac{{\rm Tr}}{2}\
 ( \   \twooneplaq \   ) \right) \le 0
\eea
Using the results of \cite{MP} one can construct a one-parameter
set of actions which have no ${\cal O}(a^2)$ and ${\cal O}(a^4)$ corrections
\begin{eqnarray}
c_1 &=&\hspace{-0.1cm}(19- 55\  c_5)/9,\ \  c_2 =  (1- 64\  c_5)/9 \nonumber\\
c_3 &=&\hspace{-0.1cm}(-64+ 640\  c_5)/45,\ \  c_4 = 1/5 - 2\  c_5
\label{coeffs}
\end{eqnarray}
Here we take $c_5=1/20$ (the same action was used in the context of improved
cooling \cite{Imprcool}). 

Following the ideas of section II, we estimate the
contribution of all loops except the plaquette stochastically.
For each loop $l\equiv \{ \mu,\nu,x\}$ of sort $2\le i\le 5$ we introduce 
the auxiliary variable $\sigma_i(l)=0,1$; and rewrite the contribution of this 
loop to Gibbs factor in the form 
\be
e^{-\frac{\beta}{2}S_{i,\mu,\nu,x}}=
\sum_{\sigma_i(l)=0,1}[\delta_{\sigma_i(l),0}
+\delta_{\sigma_i(l),1}(e^{-\frac{\beta}{2}S_{i,\mu,\nu,x}}-1)]
\ee
The resulting distribution of $\{ U,\sigma\}$ fields is used for the generation of 
independent $\{ U\}$ configurations. We shall say that for a given $\{ \sigma\}$
configuration the loop $\{l,i\}$
is ``active'' if $\sigma_i(l)=1$.

Let us describe the updating procedure in the $\{U,\sigma\}$ configuration space. 
Consider first the local updating of the
gauge fields $U$ when the $\sigma$ fields are fixed. 
The proposal value 
$U_{x,\mu}^{new}$ at a given link $\{ x,\mu\}$ is generated by heatbath 
with respect to the measure
\be
P_0[U]\;\propto \; \exp (-\frac{\beta}{2}c_1 S_{1,1}[U]) 
\label{pomeas}\ee
where $S_{1,1}$ is the plaquette action (see (\ref{act1})), 
and then accepted with probability
\be
P_{acc}=\rm{min}\Bigl( 1, \frac{P_1[U^{new}_{x,\mu},\sigma]}
{P_1[U_{x,\mu}^{old},\sigma]} \Bigr) 
\label{pacc2}\ee
where 
\be
 P_1[U,\sigma]=\prod_{\begin{array}[t]{c}(l,i)\ni \{x,\mu\} \\ \sigma_i(l)=1
 \\ \end{array}}
(e^{-\frac{\beta}{2}S_{i,l}[U]}-1
\Bigr)\label{pacc3} 
\ee
Only active loops which contain the given link $\{x,\mu\}$ contribute
to the expression in the r.h.s. of (\ref{pacc2}).

After each $N_i$ updates of fields $U$ on the entire lattice 
we update the $\sigma$ fields of sort $i$. 
For each loop $l$ we assign the values  0,1 to the 
variable $\sigma_i(l)$ with the following probabilities
\be
p_{\sigma_i(l)=0}=\exp(\frac{\beta}{2}S_{i,l}[U])\quad ;\quad 
p_{\sigma_i(l)=1}=1-\exp(\frac{\beta}{2}S_{i,l}[U])
\label{probs}\ee
Due to the absence of interaction between different $\sigma$ variables,
the probabilities (\ref{probs}) depend only on the gauge configuration, so that
$\sigma$ variables can be updated independently.

In our simulations we have measured the average values of 
$\sigma_i, \;2\le i\le 5$ which are listed in Table I. They are quite small, 
and very close to 
the perturbative estimate (\ref{klm}). This shows that one can avoid the computation
of almost all of the extended "staples" in the $U$ update. 

\vspace{.5cm}
\begin{center}
\begin{tabular}{|c|c|c|c|c|} \hline
loop & 1x2 & 1x3 & 2x2 & 3x3 \\
\hline
$\langle \sigma \rangle $& 0.0753 & 0.0199 & 0.0202 & 0.0018\\
\hline
\end{tabular}\label{table1}
\end{center}
{\bf Table I:} {\small Average value of $\sigma_i$ field
for each loop of sort $i$.}
\vspace{.5cm}

Performing numerical simulations for the 5-loop model (\ref{5Li},\ref{coeffs})
with auxiliary $\sigma$ fields, we were mainly interested in the efficiency of our 
new NMC algorithm. One can estimate the efficiency of NMC by comparing it with the 
updating procedures which are commonly used now for the simulation of multiloop 
actions like (\ref{5Li}). In the following we label these usually 
applied techniques with the collective name "Usual Monte Carlo" (UMC), 
to contrast with NMC.

We compare the computer times needed to get the same results
with NMC and UMC algorithms as follows. First, we make an analytic 
estimation of the total computational cost of one update of the $U$ fields for both 
algorithms in units of matrix (link) multiplications. Second, we extract from 
numerical simulations the integrated autocorrelation times for different observables,
in units of $U$ update. The computer time 
needed to estimate any given observable is proportional to the product 
of the computational cost per update and the autocorrelation time.   

For NMC the average computational cost of one update of the $U$ fields 
on the entire lattice is equal to
\be 
t_U^{NMC}=t_U^{pl}+4V \sum_{i=2}^{5} n_{staple}(i)*n_{mult}(i) * \langle \sigma_i\rangle 
\ee
where $t_U^{pl}$ is the cost for generating the proposal configuration with
measure (\ref{pomeas}) (i.e. the update cost for the elementary plaquette action), 
$4V$ is the number of links on the lattice, $n_{staple}(i)$ is the number
of ``staples'' which the loops of sort $i$ form for each link, $n_{mult}(i)$
is the number of matrix multiplications needed to estimate the contribution of
one staple of sort $i$, and the factor $ \langle \sigma_i\rangle $ accounts for the fact that one
needs to calculate the contribution of active loops only. One can easily check
that
\be 
n_{staple}(i)=\frac{3}{2}*P_i*s_i\quad ;\quad  n_{mult}(i)=P_i
\ee 
where $P_i\equiv 2(m_i+n_i)$ is the perimeter of loop $i$, and $s_i$ is
a symmetry factor: $s_i=1$ for square loops and $s_i=2$ for rectangular 
loops.  Then we have 
\be
t_U^{NMC}=t_U^{pl}+6V \sum_{i=2}^{5}  P_i^2  s_i  \langle \sigma_i\rangle 
\ee
On the other hand, the computational cost of one update of  
the $\sigma_i$ fields on the entire lattice is given by
\be
t_{\sigma_i}=6V s_i P_i
\label{sigcost}\ee
Here $6V s_i$ is the number of loops of a given sort on the lattice, and the perimeter
$P_i$ of the loop appears again as the number of matrix multiplications
$n_{mult}(i)$ needed to calculate the probabilities (\ref{probs}).
Since we update the $\sigma_i$ fields only once per each $N_i$ updates of the $U$
fields, the total computational cost per $U$ update
for the NMC method is
\be
t_{tot}^{NMC}= t_U^{NMC}+\sum_{i=2}^{5}\frac{t_{\sigma_i}}{N_i}=
t_U^{pl}+6V\sum_{i=2}^{5} P_i s_i (\frac{1}{N_i}+P_i  \langle \sigma_i\rangle )
\label{nmctime}\ee 

Let us note that the computational cost $t_{tot}^{UMC}$ for 
UMC of one $U$ update is approximately equal to the 
r.h.s. of expression  (\ref{nmctime}) in the limit $N_i\rightarrow\infty$ and
$\langle \sigma_i\rangle \rightarrow 1$:
\be
t_{tot}^{UMC}=t_U^{pl}+6V\sum_{i=2}^{5} P_i^2 s_i   
\label{umctime}\ee
Indeed, in the limit when all $\sigma$  are set equal to 1 and not updated
we recover the usual algorithm (certainly one should correct the expressions
(\ref{pacc2},\ref{pacc3}) for $P_{acc}$ in this case). 
 
Now we can compare the performance of our NMC algorithm with that of UMC.
The naive gain in efficiency from using NMC does not depend on the
observable measured, and is equal to the ratio between the computational costs 
(\ref{umctime}) and (\ref{nmctime}):
\be
r^{naive}_{gain}\equiv\frac{t_{tot}^{UMC}}{t_{tot}^{NMC}}=
\frac{t_U^{pl}+6V\sum_{i=2}^{5} P_i^2 s_i}
{t_U^{pl}+6V\sum_{i=2}^{5} P_i s_i (\frac{1}{N_i}+P_i  \langle \sigma_i\rangle )}
\label{naive}\ee
Now one should also take into account the increase of autocorrelation
times coming from the introduction of auxiliary variables $\sigma$ in the NMC algorithm, 
so that the real gain is
\be
r^{real}_{gain}\equiv r^{naive}_{gain} *  \frac{\tint^{UMC}}{\tint^{NMC}}
\label{gain}\ee
where $\tint^{UMC}$ and $\tint^{NMC}$ are integrated autocorrelation times for UMC
and NMC respectively. Note that $\tint^{NMC}$ is a function of the updating
frequencies $1/N_i$ of the $\sigma_i$ fields. Like $\tint$, the ratio (\ref{gain})
will also depend on the observable.

In Table II we present the autocorrelation times for averaged traces of 6 different loops
in units of $U$ updates. In the first column we show the results for the Usual Monte Carlo,
and in other columns for the NMC algorithm with different frequencies of
$\sigma$ updates for 1x2, 1x3, 2x2, 3x3 loops. In the last row we present the 
naive gain (\ref{naive}).

Let us make one useful remark. It is not necessary to keep the same updating frequencies
$1/N_i$ for all sorts $i$ of loops. Actually it is even impractical. The computational cost of
$U$ update coming from the loop of sort $i$ is proportional to the average value of 
$\sigma_i$, which is in turn proportional to the coupling (\ref{klm}). As the coupling decreases,
we should expect a reduction of the computational effort for the corresponding terms in the action.
That is not the case for the cost of $\sigma_i$ update: it does not depend on the coupling and
even increases with the nonlocality of the action term (factor $P_i$ in expression
(\ref{sigcost})). In order for the work in the $\sigma$ and in the
$U$ updates coming from loops of sort $i$ to remain comparable, one should keep the updating
frequencies $1/N_i$ proportional to $\langle \sigma_i\rangle $:
\be
\frac{1}{N_i} \;\sim  \; P_i \langle \sigma_i\rangle          
\label{var}\ee
Due to the small influence of weakly
coupled terms on the dynamics of the system, one can expect only insignificant changes in
the autocorrelation behavior as $N_i$ increases. 
These considerations are distinctly demonstrated in
Table II, where in two columns we present the results for 
updating frequencies of $\sigma$ fields varying in accordance with (\ref{var}).

 \begin{center}
\begin{tabular}{|c|c|c|c|c|c|c|c|c|c|c|} \hline
number of          & UMC    &1  &5  &5  for 1x2 & 10&10 for 1x2    &20 &30 &40 &50 \\
$U$ updates per    & no     &for&for&15 for 1x3,2x2&for&30 for 1x3,2x2&for&for&for&for \\
1 $\sigma$ update  &$\sigma$&all&all&105 for 3x3&all&210 for 3x3   &all&all&all&all \\
\hline
$\tint$ (1x1)&0.7(1)&1.9(2)&2.3(1)&2.5(2)&3.1(2)&3.2(2)&4.3(4)&4.5(4)&3.8(2)&5.1(4)\\
\hline
$\tint$ (1x2)&0.8(1)&2.6(3)&2.8(2)&3.2(2)&4.3(4)&3.9(3)&5.2(4)&5.6(5)&5.7(4)&7.7(8)\\
\hline
$\tint$ (1x3)&0.8(1)&2.7(3)&2.8(2)&3.2(2)&4.3(4)&3.9(3)&5.1(4)&5.4(5)&5.4(4)&7.4(8)\\
\hline
$\tint$ (2x2)&1.0(1)&3.4(5)&3.3(3)&3.7(3)&4.7(4)&4.7(3)&5.3(4)&5.8(6)&5.7(4)&7.5(8)\\
\hline
$\tint$ (2x3)&1.4(3)&4.2(7)&3.8(3)&4.2(3)&5.5(5)&5.7(4)&5.7(5)&6.2(6)&6.3(5)&7.7(8)\\
\hline
$\tint$ (3x3)&1.8(4)&5.0(8)&4.5(5)&5.0(5)&5.8(6)&6.4(5)&5.9(5)&6.3(6)&6.3(5)&7.5(8)\\
\hline
$r^{naive}_{gain}$&1&7.2&14.8 &18.3&17.9 &20.6&18.3&20.8&21.1&21.4\\
\hline
\end{tabular}\label{table2}
\end{center}
{\bf Table II:} {\small Integrated autocorrelation times for average loop traces in units of $U$
updates for UMC algorithm (first column) and for NMC algorithm with different frequencies of
$\sigma$ updates for 1x2, 1x3, 2x2 and 3x3 loops (other columns). The last row presents the 
naive gain for the NMC algorithm (\ref{naive}). 
}\vspace{.5cm}

Table II gives an impressive demonstration of the benefits which come from using the 
NMC algorithm. The naive gain increases substantially as we decrease the frequencies of 
$\sigma$ updates, while the 
autocorrelation times grow rather slowly. That is particularly visible 
for the runs where the updating frequencies for $\sigma$ fields are adjusted as per 
eq.(\ref{var}).
For such runs we can infer that the 'real gain' ${\cal O}(4-6)$ in computer time
(\ref{gain}) for the observables measured is large enough 
for a convincing demonstration
of the possible advantages coming from using the NMC algorithm. 

Let us make a conclusion for this section. We have applied our 
NMC algorithm for the 5-loop model (\ref{5Li},\ref{coeffs}). We have shown that
with this algorithm a significant gain in efficiency is obtained in comparison with usual
updating techniques. Finally we note that the action (\ref{5Li}) is a relatively
simple one, and one can expect a much greater gain for more complicated
highly-improved actions with many nonlocal weakly coupled terms.  

\section{Discussion}

Let us summarize our algorithm: \\
$a)$ Separate the Hamiltonian (or action) into a dominant term $c_0 W_0$, to
be calculated exactly, and correction terms $\sum_{k=1}^m c_k W_k$, to be
estimated stochastically. \\
$b)$ Shift the correction terms to guarantee $c_k W_k \le 0$. \\
$c)$ Introduce auxiliary local variables $\sigma_k(l)$, through identity
(\ref{sigpr}) (here $l$ runs through all the elementary 'bonds' which form $W_k$,
e.g. loops in gauge theory). \\
$d)$ Update the auxiliary variables $\sigma_k$ by heatbath. \\
$e)$ To update the original variables $U$, propose a new value $U'$ sampled
from the distribution $\propto e^{-c_0 W_0}$, and accept it with the Metropolis
probability 
$\rm{min}(1,\prod_{k\ge1;\sigma_k=1} (e^{-c_k W_k(U')} - 1)/(e^{-c_k W_k(U)} - 1))$.

The essential advantage of our algorithm appears in step $(e)$: only the terms
$W_k$ whose associated $\sigma_k$ is equal to 1 need to be computed. Since
on average, $\langle \sigma_k\rangle $ goes to zero with $c_k$, the computation of almost
all correction terms can be avoided.

To avoid simply shifting the cost of the algorithm to step $(d)$, we propose
to refresh the variables $\sigma_k$ infrequently, the more so as the associated
coupling $|c_k|$ gets smaller. We have pointed out that this introduction of
slow dynamics for the $\sigma_k$ does {\em not} enforce slow dynamics for the system,
since $W_k(l)$ will fluctuate regardless of the value of $\sigma_k(l)$.
Our numerical study of Section III confirms this statement.

Let us now speculate on possibilities to use our algorithm to simulate a
Hamiltonian with a very large number of terms. A specific example we have in
mind is the case of full QCD, where the measure is, for 2 flavors of Wilson
quarks:
\be
\frac{1}{Z} e^{-S_g(U)} ~~ det^2({\bf 1} - \kappa M(U))
\label{measure}
\ee
where $S_g$ is the local gauge action, $M(U)$ is a hopping matrix connecting
nearest neighbours on a $4d$ hypercubic grid, and $Z$ normalizes the distribution.
The determinant can be turned into $exp(Tr(Log({\bf 1} - \kappa M(U))))$, 
then the logarithm expanded around 1,
giving the loop expansion of the measure above:
\be
\frac{1}{Z} e^{-S_g(U) - 2 \sum_{l=4}^\infty \frac{\kappa^l}{l} Tr M(U)^l} \quad .
\label{loops}
\ee
$Tr M(U)^l$ can be represented as a sum over all closed  non-backtracking loops of 
length $l$ on the $4d$ hypercubic lattice. The number of types of contributing loops 
$n_l$ is bounded by $(2d-1)^l=7^l$, because of the branching 
factor at each hop.
Although this upper bound is not saturated, it is clear that the multiplicity
of terms of a given length $l$ grows exponentially:
\be 
 n_l \sim F_1(l) \;\alpha^l;\qquad \alpha<7
\ee
where $F_1(l)$ is a rational function of $l$, and $\alpha^l$ is the leading 
exponential ascend of the number of loops of length $l$ in the limit of large $l$.
For this reason, it seems
that sampling numerically the distribution (\ref{loops}) is a disastrous idea:
the action contains an infinite number of terms, of exponentially growing 
multiplicity. Instead, other strategies are being used, based on the 
transformation of the determinant (\ref{measure}) into a Gaussian integral.

Nevertheless, the coupling $\frac{\kappa^l}{l}$ decreases exponentially as
$l$ increases. Therefore, the auxiliary variables $\sigma_l$ associated in
our approach with various loops of length $l$ will take value 0 almost always.
From (\ref{klm}) one gets
\be 
\langle \sigma_l\rangle \; \sim c_l \sim F_2(l) \; k^l \beta^l
\ee
where the exponentially growing factor $\beta^l$ comes from the average 
trace of Dirac matrices along the loops of length $l$, $F_2(l)$ is again the 
rational function. 

If one arranges the updating frequencies for each loop $i$ as per 
eq.(\ref{var}), one can expect that the average computer time needed
for estimation of contribution of all loops of length $l$ behaves
at large $l$ as
\be
t_l \sim \; n_l l^2  \langle \sigma_l\rangle  \; \sim  F(l) * (\alpha\beta\kappa)^l
\ee
(It was pointed above that one should not expect a significant growth of
autocorrelation coming from slow dynamics for $\sigma_l$.)   
For $\;\kappa < \kappa_{ca}=\frac{1}{\alpha\beta}\;$
the computational cost $t_l$ decreases exponentially with $l$
and the total computational cost of the algorithm $t=\sum_l t_l$ 
converges to some finite value. 
Our rough estimation from fitting $\; n_l\;$ and average trace of Dirac matrices
in the interval $4\le l\le 12$ gives $\;\alpha\approx 5.4;$ $\beta\approx 1.4$,
and therefore $\kappa_{ca} \approx 0.13$. 

In the regime $\kappa <\kappa_{ca}$ we are in an interesting situation 
where the influence of very large loops
is negligible because their associated coupling in the effective action is
extremely small. Therefore, truncating the loop expansion above a certain
order will introduce a statistically unobservable bias. Equivalently, one
can freeze the associated $\sigma$ variables at the value zero, or update
them with arbitrarily low frequency. In spite of this extremely (or infinitely)
slow dynamic mode of the $\sigma$'s, the dynamics of the gauge fields are not
affected.
Note that the cost of our algorithm grows linearly with the volume $V$ of the
system. This is better than alternative approaches to the simulation of
full QCD: Hybrid Monte Carlo (cost $\propto V^{5/4}$) and MultiBoson
(cost $\propto V(Log V)^2$) \cite{Bielefeld}. In addition the stepsize, or 
typical change at each update of a gauge link $U$, does not seem restricted
a priori for small quark mass, unlike in the two alternative approaches above.
Unfortunately the possible high efficiency of our algorithm is 
counterweighted by its extreme programming complexity.

A less speculative use of our algorithm for full QCD consists of truncating
the loop expansion eq.(\ref{loops}) to some order $l_{max}$, and to represent
the higher orders with the MultiBoson approach \cite{MB}. This strategy,
called ``UV-filtered MultiBoson'', has already been used successfully
\cite{Forcrand}. However, in Ref.\cite{Forcrand} the loop expansion is truncated 
to its lowest term $l=4$, because the exact evaluation of larger loops is too
time-consuming. With our stochastic approach, these larger loops can be 
estimated at low cost. We expect this composite strategy to be particularly
efficient.

\vspace{.5cm}

{\bf Acknowledgments:}
We acknowledge communications with T. DeGrand, M. Ilgenfritz and U. Wenger. 
T.B. was supported by
INTAS under grant 96-0370 and Russian Basic Research Fund 
under grant 99-01-00190.

\end{document}